\documentclass[final,5p,times,twocolumn]{elsarticle}

\usepackage{amssymb}

\journal{Astroparticle Physics}

\begin{document}

\begin{frontmatter}



\title{Measurement of the near-infrared fluorescence of the air for the detection of ultra-high-energy cosmic rays}


\author[INFNPD]{E. Conti\corref{c1}}
\author[UNIPD,INFNPD]{G. Sartori}
\author[UNIPD,INFNPD]{G. Viola}

\address[INFNPD]{INFN sez. di Padova, Via F.Marzolo 8, 35131 Padova, Italy}
\address[UNIPD]{Dip. di Fisica ``G.Galilei", Universit\'a di Padova, Via F.Marzolo 8, 35131 Padova, Italy}

\cortext[c1]{Corresponding author. email: enrico.conti@pd.infn.it, phone: ++39.049.8277198, fax: ++39.049.8277102}

\begin{abstract}

We have investigated the fluorescence emission in the Near Infrared from the air and its main components, nitrogen and oxygen. The gas was excited by a $95~kV$ electron beam and the fluorescence light detected by an InGaAs photodiode, sensitive down to about $1700~nm$. We have recorded the emission spectra by means of a Fourier Transform Infrared spectrometer. The light yield was also measured by comparing the Near Infrared signal with the known Ultraviolet fluorescence, detected by a Si photodiode.

The possibility of using the Near Infrared fluorescence of the atmosphere to detect Ultra-High-Energy Cosmic Rays is discussed, showing the pros and the cons of this novel method.

\end{abstract}


\end{frontmatter}

\newpage
\section{Introduction}

The fluorescence of the nitrogen gas in the atmosphere in the visible (VIS) and ultraviolet (UV) regions was independently proposed in the sixties of last century 
by Greisen \cite{greisen}, Delvaille et al. \cite{delvaille}, Suga \cite{suga}, and Chudakov \cite{chudakov} as a way to detect high energy showers created by cosmic rays impinging on the earth atmosphere.  Nowadays the air fluorescence method is a well established technique used by the experiments HiRes \cite{hireswww}, which is the continuation of the pioneering Fly's Eye experiment \cite{flyseye}, AUGER \cite{augerwww}, and by the Telescope Array project \cite{TAwww}.

The main emission of the fluorescence light is in the UV region between $300$ and $450~nm$.
The transparency of the atmosphere at those wavelength is limited by the presence of molecular oxygen and ozone, which absorb the UV photons.  The Rayleigh scattering, whose cross section scales as $1/\lambda^4$, and the Mie scattering further reduce the number of photons which arrive at the detector. We introduce an extinction length $\Lambda(\lambda)$, function of the wavelength $\lambda$, which enters in the transmission of the light through the relationship
$
 I(x,\lambda) = I_0\cdot exp(-x/\Lambda(\lambda))
$, where $I(x,\lambda)$ is the intensity of a collimated light beam of wavelength $\lambda$ after having travelled for the distance $x$ in the atmosphere, and $I_0$ is the initial intensity (at $x=0$). For the UV,  $\Lambda \lesssim 10~km$.

This fact poses two main drawbacks. First, the detector cannot be placed at sea level, but at altitude. For example, AUGER is at about $1400~m$ above sea level, HiRes at $1700~m$. 

Second, the detection range of the instruments extends to a distance of the order of $\Lambda$.  AUGER, for example, is capable to detect the highest energy air showers at a distance of the order of $30~km$ \cite{AUGER2010}. This is correlated to the total observation rate, which goes roughly as $\Lambda^2$, which is therefore limited also by the atmospheric transmission.

The atmosphere is very transparent in some regions of the near-infrared (NIR) region, below $3~\mu m$.  The molecules which present absorption bands are water and, in a minor extent, CO$_2$. Several high transparency windows are present: from 0.8 to $0.9~\mu m$, from 0.95 to $1.05~\mu m$, from 1.15 to $1.3~\mu m$,  from 1.5 to $1.8~\mu m$, from 2.0 to $2.4~\mu m$ (see, for example, ref.\cite{manducabell}). Moreover, the Rayleigh scattering is negligible in the NIR,  because of the $1/\lambda^4$ dependence.
If the air fluoresces at those wavelengths, then there is the possibility to detect high energy cosmic rays impinging on the atmosphere at a very long distance from the detector, thus increasing the observation rate.

The spectroscopy of the molecule of nitrogen is rather complex and has been extensively studied in the last century. An exhaustive review can be found in ref.\cite{n2spectroscopy}. Despite of this, to our knowledge there are no investigations on the infrared fluorescence emission from N$_2$ or air excited by ionizing radiation. The only research which extends somewhere in the NIR region was performed in the sixties by Davidson and O'Neil \cite{davidson}, who used photomultipliers to reach the maximum wavelength of $1050~nm$.

In this paper we present our first measurements of the fluorescence light in the NIR region of air, nitrogen, and oxygen. The light emitted by the gas, excited by a $95~kV$ electron beam, was detected by an InGaAs photodiodes, sensitive down to about $1.7~\mu m$. Spectra have been recorded at atmospheric pressure and room temperature by mean of a Fourier Transform Infrared Spectrometer. We also report on the light yield, which has been obtained by comparing the NIR signal with the known fluorescence output in the UV region, measured by means of a silicon photodiode.

\section{Experimental details}

\begin{figure*}[!t]
\centering
\includegraphics[width=6.0in]{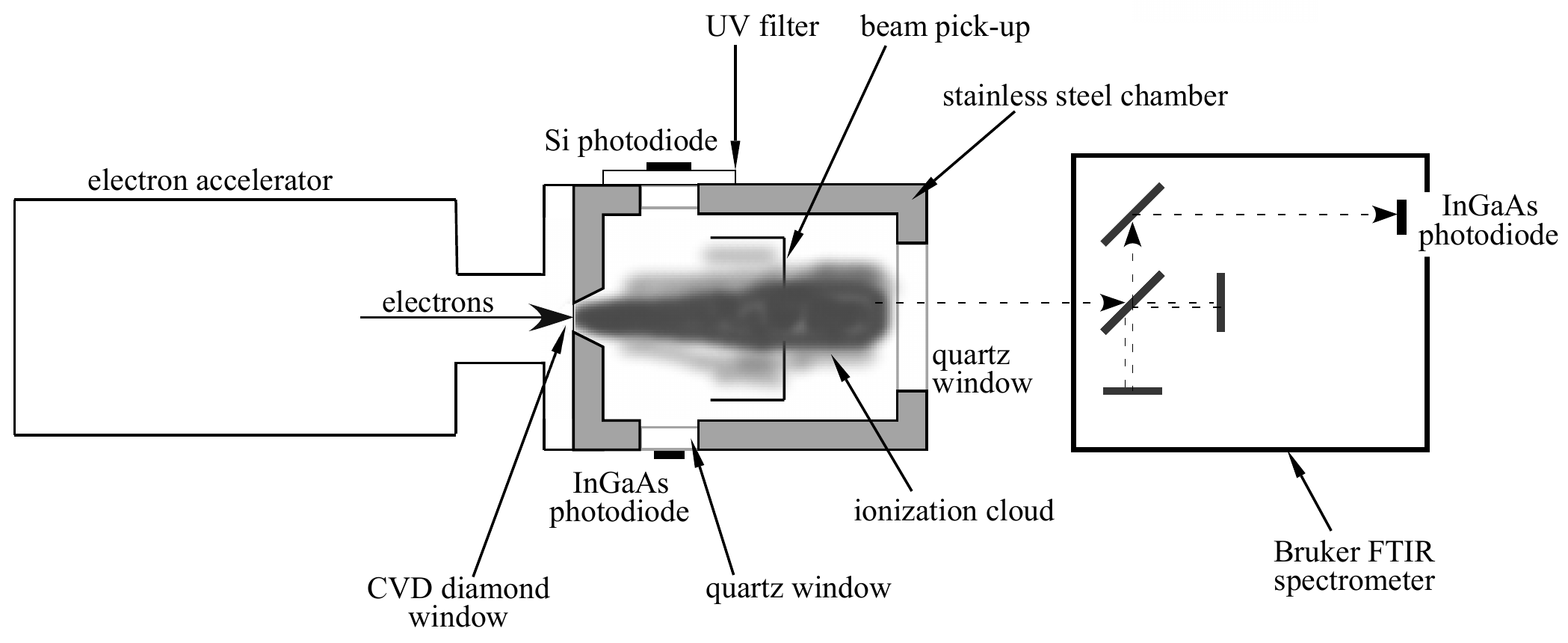}
\caption{A schematic view of the experimental setup (not to scale). Not shown are the vacuum and the gas lines.}
\label{fg:setup}
\end{figure*}

A schematic of the experimental setup is shown in fig.\ref{fg:setup}.

An electrostatic gun \cite{kimball} accelerated electrons, emitted by a hot filament, up to $100~keV$ kinetic energy. Electrons exited the gun through a CVD diamond window \cite{diamondmaterials}, $20~\mu m$ thick, whose external side had been metalized to screen the fluorescence light emitted by the diamond itself.
The gun was operated at  $95~kV$ in pulsed mode, with pulses $300~\mu s$ long and rate of $50~Hz$. In those conditions,  the electron current behind the diamond windows was a few hundreds $\mu A$.

The chamber containing the gas was a stainless-steel cylinder with the axis along the beam path and volume about $300~cm^3$. The chamber size was a compromise between the range of the electrons in the gas and the need of an adequate solid angle for the detection of the light. The ionization cloud was almost entirely contained in the chamber, and the electrons that hit the materials of the chambers did not affected neither disturbed the measurements, as proved in section 3.
Perpendicular to the axis, one opposed to the other, two quartz windows allowed to view inside the chamber. They were $10~mm$ thick in order to stop the X-rays generated in the chamber, that could alter the signals of the photodiodes. By construction, they looked exactly at the same gas region, a small fraction (about $5\%$) of the chamber volume. The light to the spectrometer exited through a third quartz window on the closing flange of the vessel. Inside the chamber, an electrode collected the ionization charge for the beam monitoring and normalization, which was fed into a charge amplifier. The output signal $Q_{beam}$ was displayed on a digital oscilloscope and also used as a feedback to stabilize the electron gun current against long term drifts. $Q_{beam}$ was compared to a reference voltage level, and the difference signal was used to opportunely bias the grid in front of the filament, which regulated the electron beam current intensity before acceleration. 

Gas from certified bottles was flown in the chamber at room temperature and atmospheric pressure with a flux of $0.2~\ell/min$.  We used N$_2$ (impurity concentration $\le 5.5~ppm$), O$_2$ (impurity concentration $\le 5~ppm$), and dry air, which is a mixture of $80\%$ nitrogen and $20\%$ oxygen (impurity concentration $\le 10~ppm$).

\begin{figure*}[!t]
\centering
\includegraphics[width=5.0in]{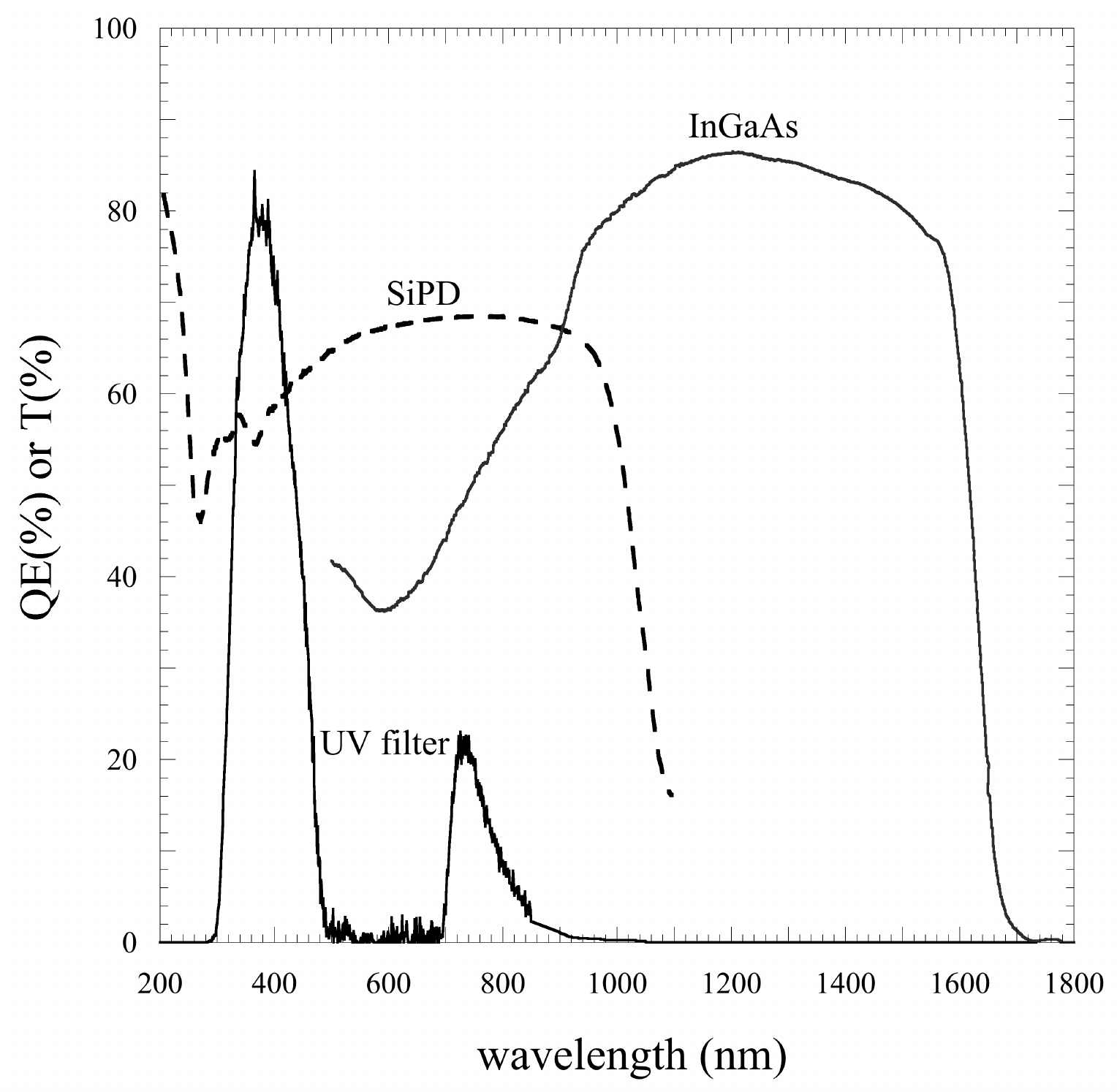}
\caption{Quantum efficiency curves of the PDs. Dashed line = Si PD; full line = InGaAs PD. It is also shown the total transmission curve of the colored filters used to isolate the UV region.}
\label{fg:PD_QE}
\end{figure*}

The fluorescence light was detected by two solid state photodiodes (PDs), which could be placed on the two lateral windows of the chamber or inside the spectrometer.  The UV light was detected by a squared Si PD \cite{sipd}, $10\times 10~mm^2$ area, with a quantum efficiency (QE) of about $55\%$ from $300$ to $400~nm$. The infrared radiations was detected by a round InGaAs PD \cite{ingaaspd}, $5~mm$ diameter, sensitive from about $500~nm$ to $1.7~\mu m$ and cooled down to about $-40~^\circ C$ by a built-in Peltier cooler. Its QE is above $80\%$ from $1.1$ to $1.5~\mu m$.  The QE curves of both PDs are shown in fig.\ref{fg:PD_QE}. 
Each photodiode was readout by a dedicated charge preamplifier, based on the UA1 hybrid \cite{QAUA1}. The gain was measured by injecting a known calibration charge. 
The output signals  were displayed and measured by means of a digital oscilloscope.

Since the Si PD is also sensitive to the NIR radiation, we used some colored glass filters to select only the UV photons. The combinations of two filters, the filter FGB25 (by Thorlabs) and the filter HA30 (by Hoya Corp.), produced the transmission curve  shown in fig.\ref{fg:PD_QE}. Only two bands are allowed, from $300$ to $500~nm$, which is the region of the UV fluorescence, and from $700$ to $900~nm$, where the fluorescence light output of air and nitrogen is negligible \cite{davidson}.

Spectra were recorded and analyzed by a Fourier Transform Infrared (FTIR) spectrometer (mod. Equinox55 by Bruker Optics) which is essentially a two-arm Michelson interferometer with a $632~nm$ He-Ne laser as reference light source. 
For the Equinox55  the minimum measurable wavelength was $632~nm$.
For details of this well known spectroscopic technique see for example ref.\cite{chamberlain,griffith}.  The spectrometer was controlled and the data were elaborated by the software OPUS 5.5 (by Bruker Optics). 

The laser light covered the same path of the radiation under measurement, and reached the detector, producing a very large noise. To filter the He-Ne light, we placed a colored glass filter in front of the photodiode, which transmitted only photons with wavelength longer than $780~nm$ (filter FGL780 by Thorlabs).

The FTIR spectrometer recorded the interferogram of the light source as a function of the wavenumber $k = 1/\lambda$. 
In the calculation of the Fourier transform of the interferogram to obtain the spectrum, we  applied the apodization function Blackman-Harris (3-terms), and the phase correction with the Mertz method (no peak search) \cite{chamberlain,griffith}. 

The spectrometer resolution on $k$ ranged between 20 and $5~cm^{-1}$ but the final spectrum resolution was worse because of the low spatial coherence of the light source and of the mathematical operations involved in the interferogram-to-spectrum conversion.
Note that a constant resolution $\delta k$ on $k$ implies a non constant wavelength resolution $\delta \lambda$, given by $\delta \lambda = \lambda^2\cdot\delta k$.

To reduce fluctuations and noise in the spectrum, the final result was obtained by 
averaging together many different spectra (between 10 and 20).

\section{Results}

As preliminary operation, we verified that no light emission occurred from the CVD diamond window or from the quartz windows or other materials of the chamber. To accomplish that, we evacuated the chamber with a rotative pump and run the electron gun at the highest beam current and energy. We could not observe any light signal.

\subsection{Spectra}

\begin{figure*}[!t]
\centering
\includegraphics[width=5.0in]{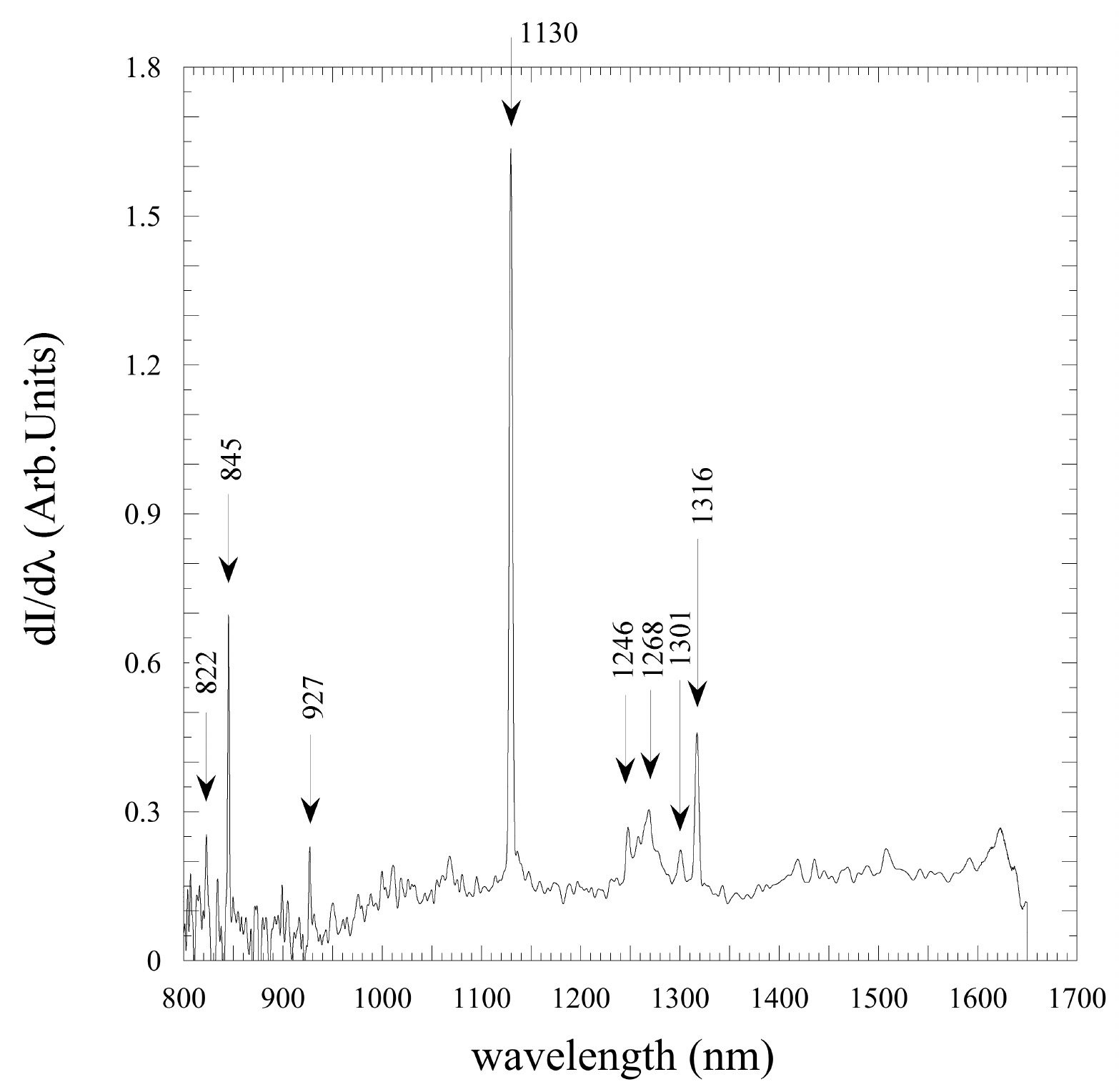}
\caption{Fluorescence spectrum of oxygen at room temperature and atmospheric pressure. Arrows show the main lines reported in table \ref{tab:o2lines}.}
\label{fg:spettroo2}
\end{figure*}

Figures \ref{fg:spettron2}, \ref{fg:spettroo2}, and \ref{fg:spettrodryair} show the measured spectra of nitrogen, oxygen and dry air, respectively. The spectra were recorded at atmospheric pressure and room temperature ($298~K$), and measured with a resolution $\delta k = 5~cm^{-1}$, apart from oxygen. They are corrected for the transmission curve of the filter  FGL780 and for the QE curve of the InGaAs PD, and truncated to $1650~nm$. To identify the atomic transitions, we used the NIST Atomic Spectra Database \cite{nist}.

\subsubsection{Nitrogen}

Molecular nitrogen has a very rich spectrum, with numerous band systems corresponding to electronic transitions of N$_2$ and N$_2^+$. The near infrared region is dominated by the First Positive System ($B^3\Pi_g - A^3\Sigma_u^+$) (FPS). A less important system is the $A^2\Pi_u - X^2\Sigma_g^+$ Meinel System (MS) of the ionized nitrogen. The Herman Infrared System (HIS) also appears, which extends from $700$ to $910~nm$ \cite{n2spectroscopy}.

\begin{figure*}[!t]
\centering
\includegraphics[width=5.0in]{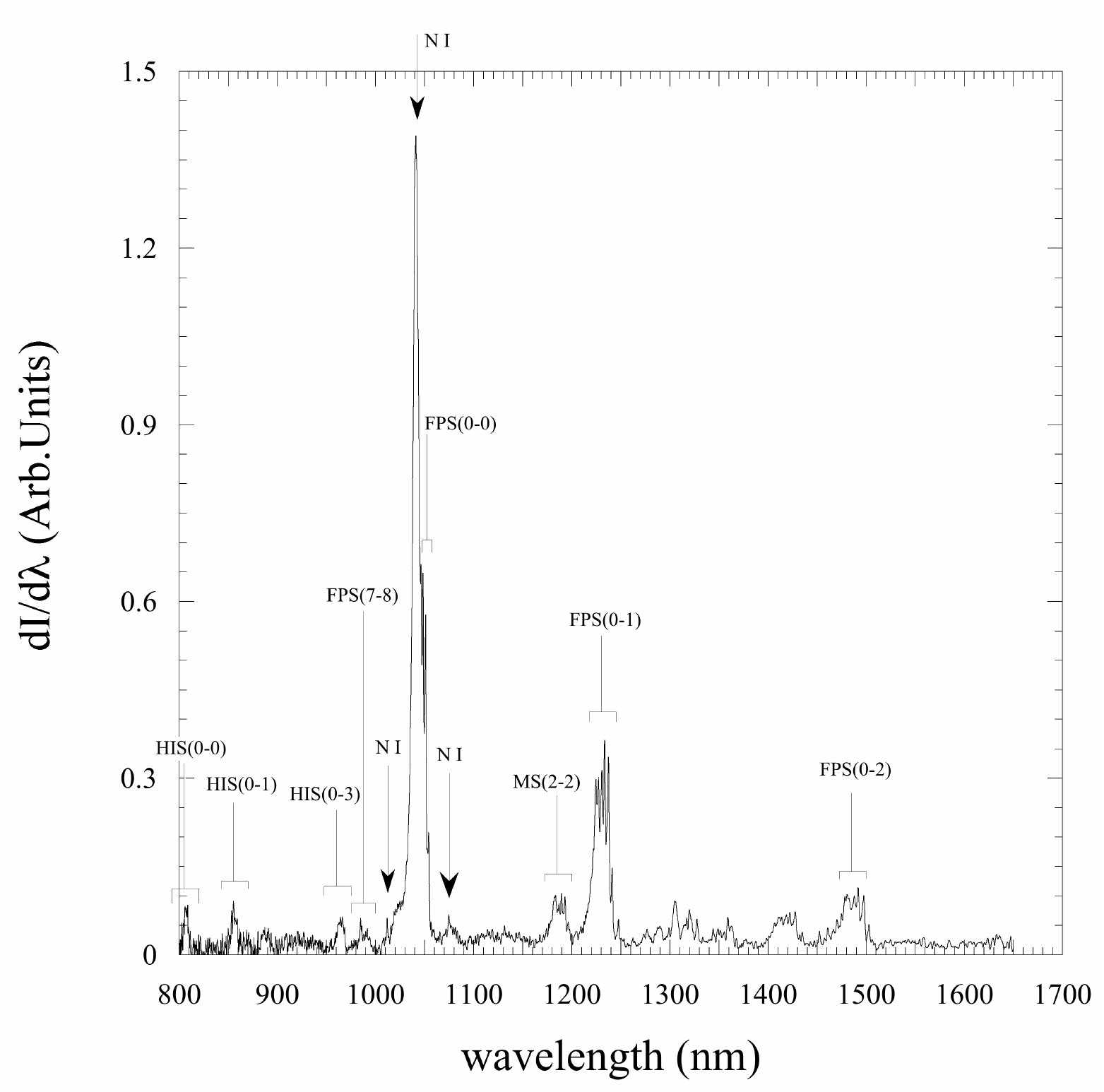}
\caption{Fluorescence spectrum of nitrogen at room temperature and atmospheric pressure. 
The bands and lines clearly identified are shown: HIS = N$_2$ Herman infrared system; N I = atomic neutral atom; FPS = N$_2$ first positive system; MS = $N_2^+$ Meinel system.}
\label{fg:spettron2}
\end{figure*}

\begin{figure*}[!t]
\centering
\includegraphics[width=5.0in]{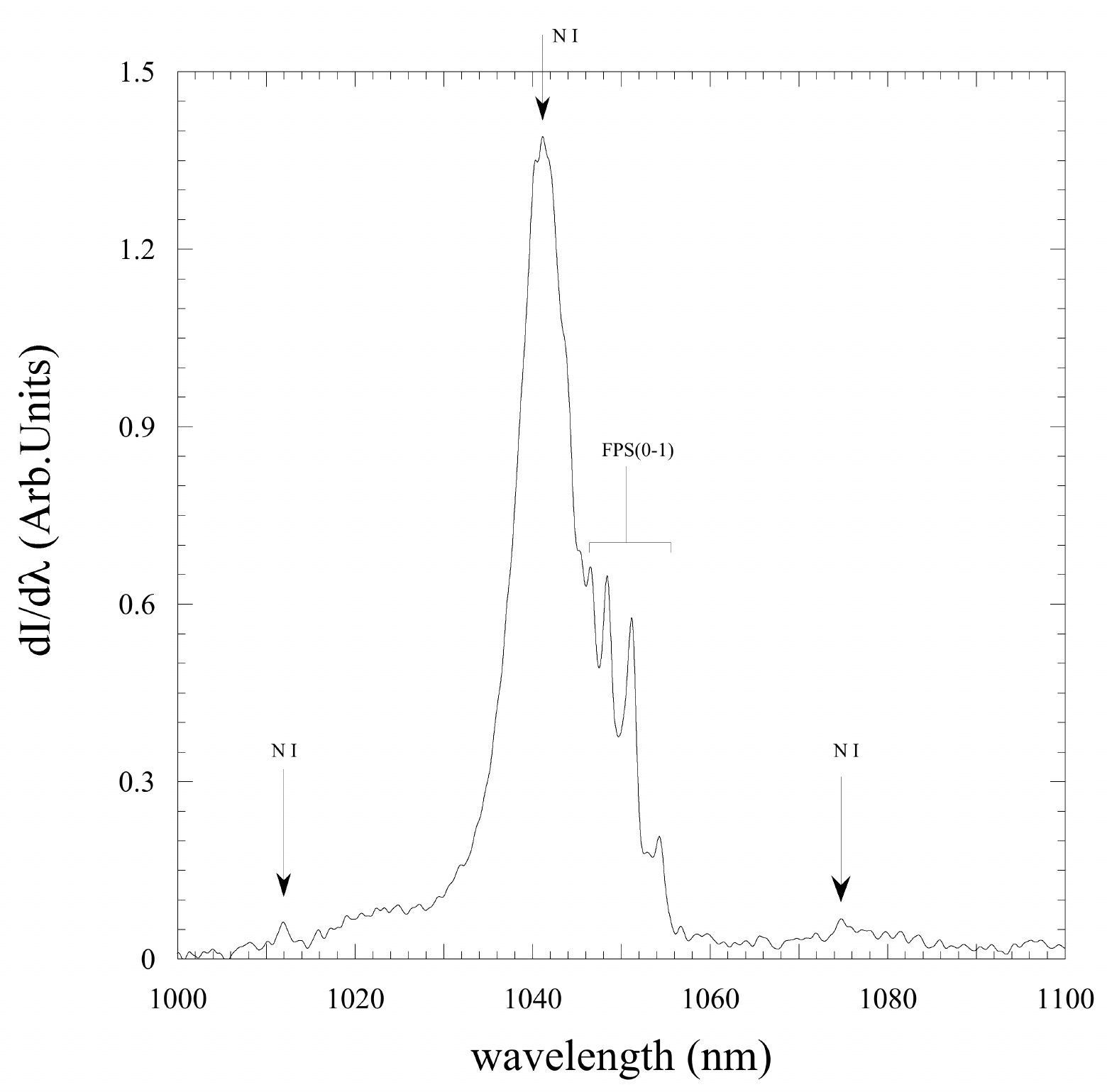}
\caption{Expanded view of the fluorescence spectrum of nitrogen at room temperature and atmospheric pressure in the wavelength region $1000\div1100~nm$. The spectrum is characterized by the overlap of two nearby atomic transitions at $1040~nm$ (not resolved), and of the FPS (0-0) band.}
\label{fg:spettron2zoom}
\end{figure*}

The spectrum in fig.\ref{fg:spettron2} is dominated by the unresolved band around $1040\div1050~nm$, which has been already observed by \cite{davidson}. The band is the overlap of two nearby atomic transitions at $1040~nm$, characteristic of the forbidden doublet transition $[N~I]_{32}$, and of the FPS (0-0) band (fig.\ref{fg:spettron2zoom}). The second more intense band is the (0-1) band of the FPS centered at $1230~nm$. Other minor bumps has been identified as belonging to the HIR, to the MS (at about $1187~nm$) and to the FPS (the (0-2) band centered at $1485~nm$). Two very weak atomic lines of the neutral nitrogen atom are distinguishable at the side of the main band, at $1011~nm$ (transition $3d~^4F \to 3p~^4D^0$) and at $1076~nm$ (transition $3d~^4P \to 3p~^4P^0$).

Because of the weak intensity and the poor resolution, bands from about 1250 to $1450~nm$ cannot be unambiguously identified.

\subsubsection{Oxygen}

Since the oxygen fluorescence light output is sensibly lower than the nitrogen or dry air one, we run the spectrometer with a resolution $\delta k = 20~cm^{-1}$.

Molecular oxygen is a weak light emitter since transitions from excited states to the ground state are strongly forbidden, for most of the excited states. As a fact, the only line we could clearly identify to belong to a molecular transition is the faint line at $1268~nm$, which is the well-known $^1\Delta_g \to ^3\Sigma_g$ magnetic dipole intercombination transition \cite{krupenie_o2}. All other lines have been identified as atomic transitions from the excited neutral oxygen (O I), created by electron-impact dissociation of the O$_2$ molecule, or from the single ionized atom (O II).  

The list of the lines and relative transitions is reported in table \ref{tab:o2lines}. The line at $1130~nm$ is the sum of two very close lines, which were not resolved by the spectrometer. The line intensities, after the continuum subtraction, are normalized to the highest intensity line.

\begin{table*}[hp]
\caption{Observed line in the oxygen spectrum. O I is the neutral oxygen atom, O II is the single ionized atom. For the calculation of the intensity, the continuum has been subtracted.}
\label{tab:o2lines}
  \begin{center}
   \begin{tabular}{ c c c c } \hline \hline
   wavelength (nm) & transition & species & intensity \\ \hline \hline
   822 & $3p^{'} ~ ^3D_{3,2} \to 3s^{'} ~^3D^0_3$ & O I & 10 \\
   845 &  $3p~ ^3P  \to 3s~^3S^0$ & O I & 24  \\ 
   927 & $3d^5~D^0 \to 3p~^5P$ & O I & 6 \\
   1130 & $3d~^3D^0 \to 3p~^3P$ (1129~nm)  & O I & 100 \\
            & $4s~^5S^0 \to 3p~^5P$ (1130~nm)  &       & \\
   1246 & $5f~^5F \to 3d~^5D^0$ & O I & 5 \\
   1268 & $^1\Delta_g \to ^3\Sigma_g$ & O$_2$ & 7 \\
   1301 & $4p~^2D^0 \to 4s~^2P$ & O II & 13  \\
   1316 & $4s~^3S^0 \to 3p~^3P$ & O I & 29 \\
 \hline \hline
\end{tabular}
\end{center}
\end{table*}

\subsubsection{Dry air}

\begin{figure*}[!t]
\centering
\includegraphics[width=5.0in]{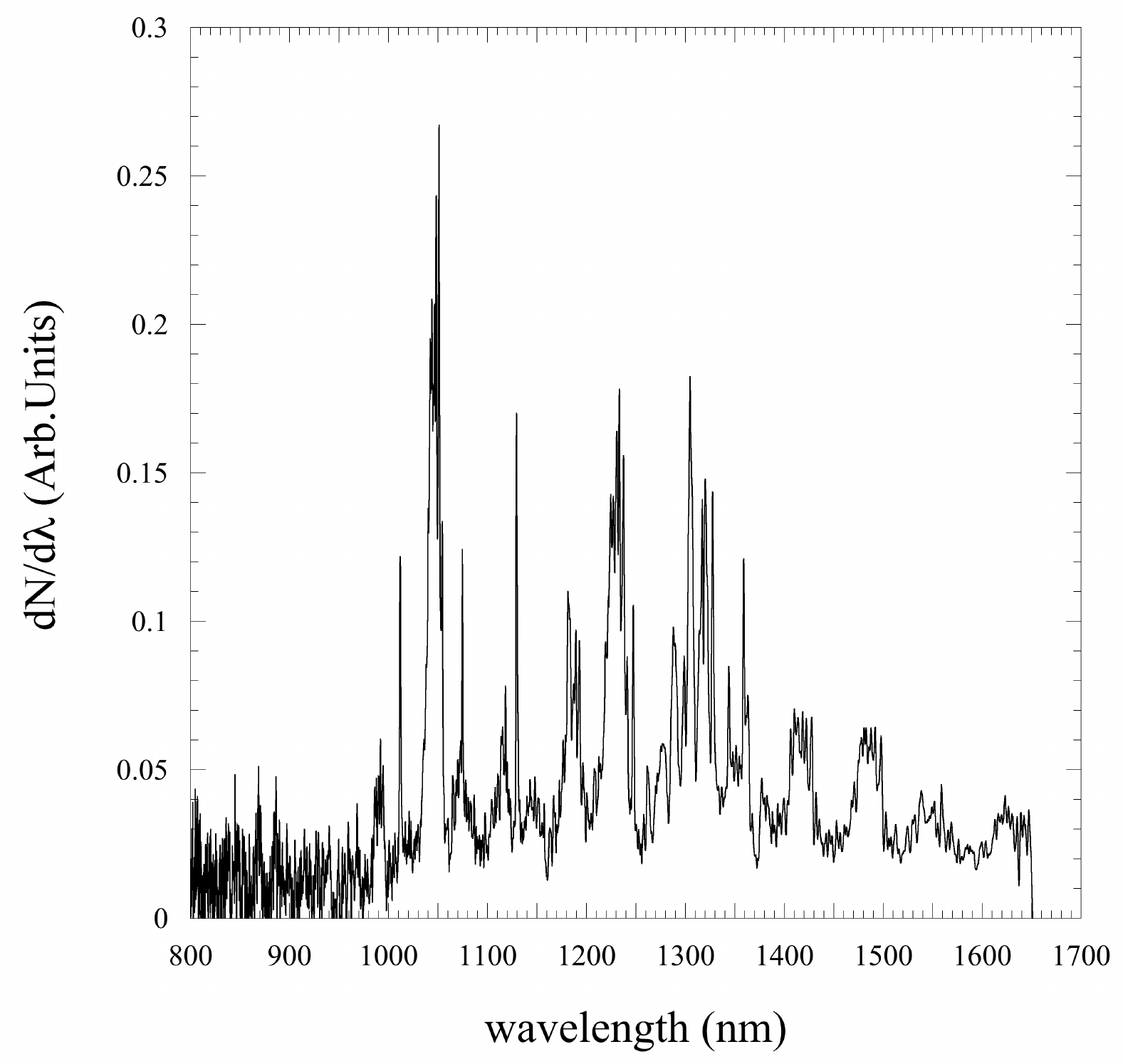}
\caption{Fluorescence spectrum of dry air at room temperature and atmospheric pressure.}
\label{fg:spettrodryair}
\end{figure*}

\begin{figure*}
\centering
\includegraphics[width=5.0in]{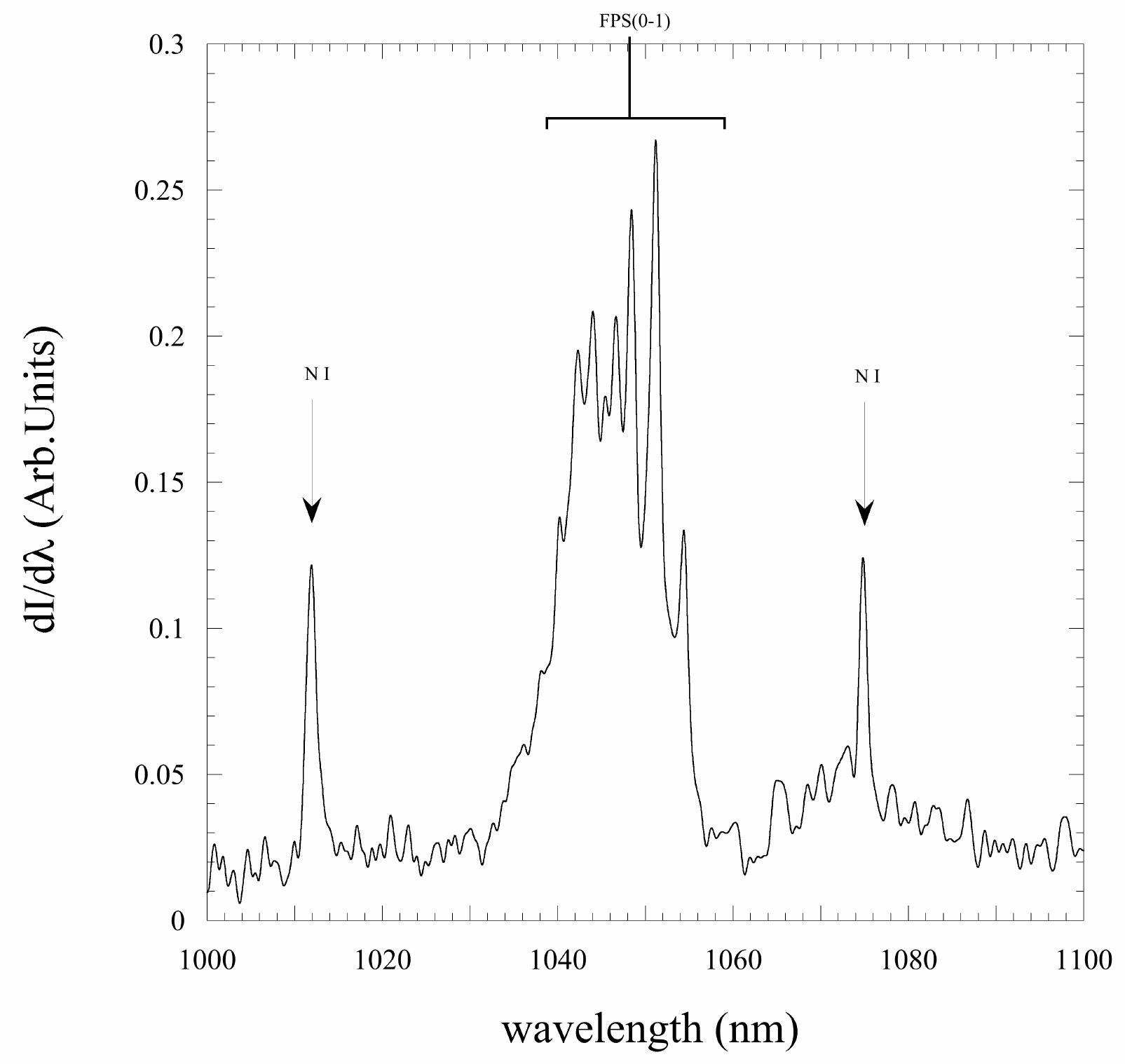}
\caption{Expanded view of the fluorescence spectrum of dry air at room temperature and atmospheric pressure in the wavelength region $1000 \div 1100~nm$. The spectrum is characterized by the  FPS (0-0) band and by two atomic transitions  of the neutral nitrogen.}
\label{fg:spettrodryairzoom}
\end{figure*}

From the comparison with the spectra of  N$_2$ and O$_2$ (fig.\ref{fg:spettron2} and fig.\ref{fg:spettrodryair} and also the expanded views fig.\ref{fg:spettron2zoom} and fig.\ref{fg:spettrodryairzoom}), it is evident that the spectrum of the dry air is not simply the weighted sum of the two. Atomic and molecular interactions among different elements change the population of the excited states of the species and the de-excitation spectrum.
The spectrum of dry air is densely populated with bumps and peaks, which could not always be easily identified. It is dominated by the nitrogen FPS band (0-0) at $1046~nm$, but, contrary to pure N$_2$, the doublet atomic lines at $1040~nm$ are absent (compare fig.\ref{fg:spettron2zoom} and fig.\ref{fg:spettrodryairzoom}). Many atomic lines of the excited neutral nitrogen atom are present, some of which were not observed in the pure gas, or have a different intensity, such as, for example, the two lines at $1100~nm$ and $1076~nm$ at the sides of the FPS(0-0) band (fig.\ref{fg:spettron2zoom} and \ref{fg:spettrodryairzoom}). The only lines ascribable to oxygen are the O I lines at $845$ and $1130~nm$. Many bumps are probably bands of the N$_2$ FPS. Since an exhaustive discussion of all the features of the spectrum is beyond the purpose of this paper, we do not go into further detail. In table \ref{tab:dryair} we report the list of the main peaks together with their associated transition (when identified), and their intensity.

\begin{table*}
\caption{Observed peaks in the dry air spectrum. When identified, the transition and the atomic/molecular species are reported. For the calculation of the intensity, the continuum has been subtracted.}
\label{tab:dryair}
  \begin{center}
   \begin{tabular}{c c c c c} \hline \hline
   wavelength (nm) & transition & species & intensity  \\ \hline \hline
   845 &  & O I &   1 \\
   869 &  & N I &  2 \\
   886 &  &      &  1  \\
     991 &  FPS (2-2)  & N$_2$ & 13 \\
   1012 &    & N I & 5 \\ 
   1046 &   FPS (0-0) &  N$_2$     & 97  \\
   1075 &                    & N I & 5 \\
   1116 &  &      & 12  \\
   1118 &  & N I & 1 \\
   1130 &  & O I & 9  \\
   1186 &  MS & N$_2^+$      & 43  \\
   1230 &  FPS (0-1) &   N$_2$     & 100  \\
   1247 &  & N I  & 4 \\
   1262 &  &        & 3 \\
   1276 &  &        & 9  \\
   1289 &  &        & 15 \\
   1298 &  &        & 5 \\
   1305 &  &        & 30 \\
   1317 &  &        & 3 \\
   1321 &  &              & 19  \\
   1328 &  &              & 9 \\
   1344 &  & N I         & 3  \\
   1359 &  &  N I        & 6  \\
   1363 &  &  N I        & 5   \\
   1416 &  &              & 35  \\      
   1484 &  FPS (0-2) &  N$_2$      & 47 \\
   1545 &  &             & 27  \\
   1623 & &             & 14 \\
 \hline \hline
\end{tabular}
\end{center}
\end{table*}

\subsection{Light yield}

\begin{figure*}[!t]
\centering
\includegraphics[width=5.0in]{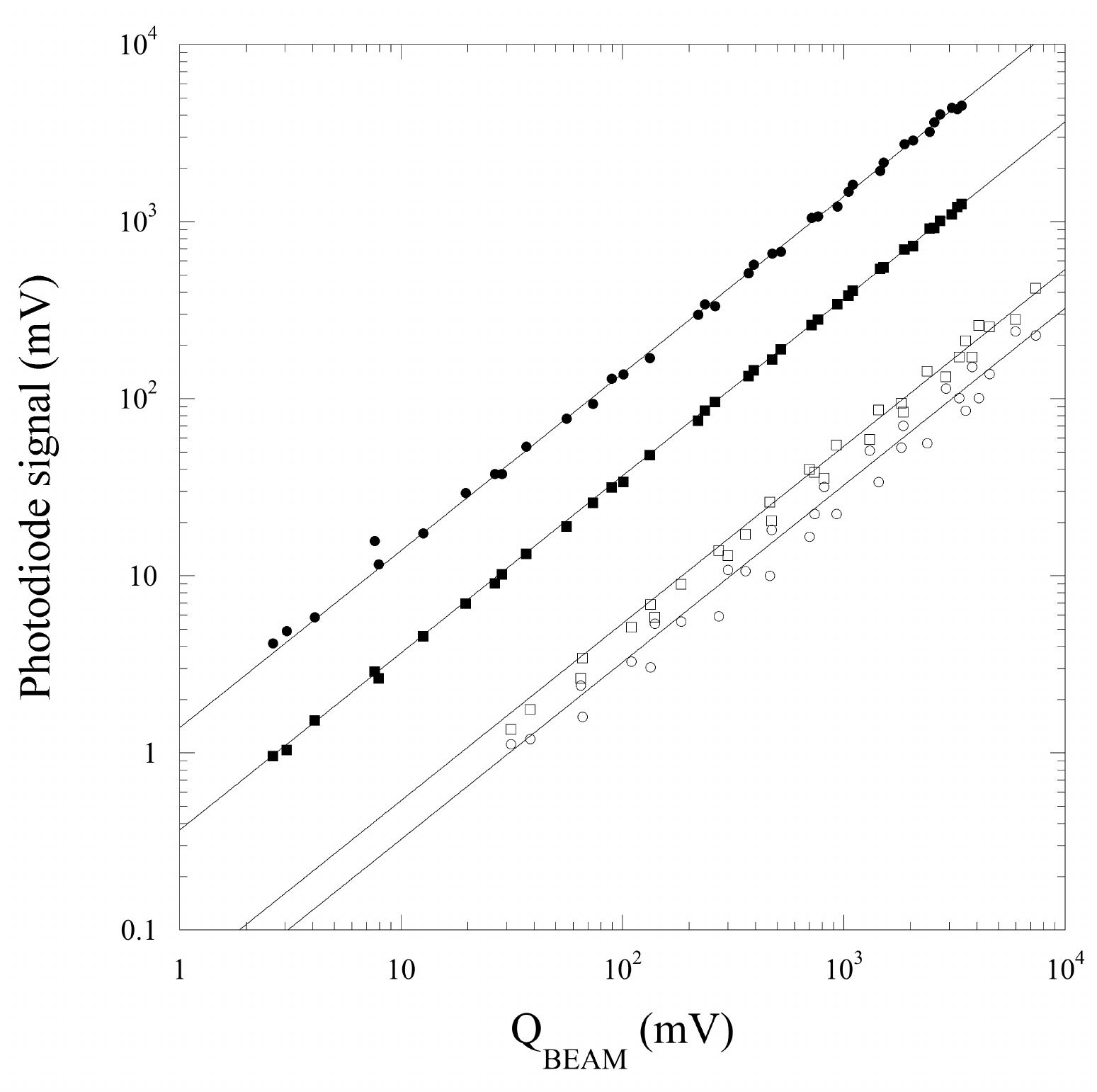}
\caption{Dependence of the light signal  of the Si and InGaAs PDs, $V_{UV}$ and $V_{IR}$ respectively, upon the charge collected by the beam pickup. Circles = Si PD data; squares = InGaAs PD data; Full symbols = total PD signals; open symbols = X-ray contribution. The full lines are the least square fits to the data with the $y = \alpha x$ function.}
\label{fg:yield}
\end{figure*}

To measure the light yield $Y_{IR}$ of the dry air in the NIR region we compared the signal output of the Si and InGaAs PDs, $V_{UV}$ and $V_{IR}$, respectively.  
Preliminarily, the symmetry of the apparatus was verified by exchanging the two detectors and checking that their signal amplitude did not change.
The data were taken at different beam intensities to verify whether saturation effects or other effects due to the high ionization density of the gas, are present. The beam intensity was measured through the charge $Q_{beam}$ collected by the beam pickup, which is proportional to the energy deposited in the gas. Each point was the average of about a hundred signals, performed by the digital oscilloscope. As can be seen from fig.\ref{fg:yield}, the dependence of $V_{UV}$ and $V_{IR}$ on $Q_{beam}$ is proportional to the charge over 3 orders of magnitude of beam intensity, which consequently excludes non-linear effects due to the ionization density. The photodiode signals had a small contribution from the X-rays generated inside the chamber and not absorbed by the quartz windows. To quantify the effect, we placed in front of the photodiodes a thin black paper foil, which is opaque to the NIR light but completely transparent to the X-rays, and measured the signal vs. $Q_{beam}$.  The X-ray signal (fig.\ref{fg:yield}) is proportional to $Q_{beam}$ in the whole explored range. The effect is however small, being about $2\%$ for SiPD and $17\%$ for InGaAs.
We can therefore write

$$
V_\alpha (Q_{beam}) = A_\alpha \cdot Q_{beam} + B_\alpha  \cdot Q_{beam}~~~~~~~~(\alpha = UV, IR)
$$

\noindent
where the $B_\alpha$ coefficients are related to the X-ray signals and are obtained from the fits of fig.\ref{fg:yield}. The  $A_\alpha$ coefficients are related to the total photon yield $Y_{IR}$ and $Y_{UV}$ in the near-infrared and in the UV region through the relationships:

\begin{eqnarray}
    \label{eq:1}
A_{UV} = C \cdot Y_{UV} \cdot \Omega_{UV} \cdot G_{SiPD} \cdot \overline{\epsilon_{UV}}
\end{eqnarray}
\begin{eqnarray}
    \label{eq:2}
A_{IR} = C \cdot Y_{IR} \cdot \Omega_{IR} \cdot G_{InGaAs} \cdot \overline{\epsilon_{IR}}
 \end{eqnarray}
 
\noindent
where $C$ is a constant,  $\Omega_{UV}$ and $\Omega_{IR}$ are the solid angles for the Si and InGaAs detectors, respectively, $G_{SiPD}$ and $G_{InGaAs}$ the gains of the two charge amplifiers. $\overline{\epsilon_{\alpha}}$  is the average quantum efficiency $\epsilon_{\alpha}(\lambda)$ of the photodiode (fig.\ref{fg:PD_QE}) weighted by the appropriate light fluorescence spectrum $S_{\alpha}(\lambda)$, which must also include the transmission curve $T_{\alpha}(\lambda)$ of the optical filter when present:

$$ 
\overline{\epsilon_{\alpha}} = {\int \epsilon_{\alpha}(\lambda) \cdot T_{\alpha}(\lambda) \cdot S_{\alpha}(\lambda)  d\lambda \over \int S_{\alpha}(\lambda)d\lambda } ~~~~~~(\alpha=UV, IR)
$$

\noindent
with $T_{IR}(\lambda)=1$ and $T_{UV}(\lambda)$ given in fig.\ref{fg:PD_QE}. $S_{IR}(\lambda)$ is the experimental spectrum of fig.\ref{fg:spettrodryair} while $S_{UV}(\lambda)$ has been taken from \cite{spettrouv}. 

From the ratio of eq.(\ref{eq:1}) and (\ref{eq:2}), we obtain  $Y_{IR}$ as:

$$
Y_{IR} = Y_{UV} \cdot {A_{IR} \over A_{UV}} \cdot {\Omega_{UV} \over \Omega_{IR}} \cdot {G_{SiPD} \over G_{InGaAs}} \cdot {\overline{\epsilon_{UV}} \over \overline{\epsilon_{IR}}} 
$$

\noindent
The quantities $A_{IR}$ and $A_{UV}$ have been obtained from the linear fits of fig.\ref{fg:yield}, after X-ray background subtraction. The result is ${A_{IR} \over A_{UV}} = 0.231 \pm 0.002$.  Since:

${\Omega_{UV} \over \Omega_{IR}} = 4.19 \pm 0.15 $

${G_{SiPD} \over G_{InGaAs}}  = 0.500 \pm 0.006$

$\overline{\epsilon_{UV}} = 0.35$

$\overline{\epsilon_{IR}} = 0.80$
 
 \noindent
the ratio of the light yields is:

 $$
 \frac {Y_{IR}}{Y_{UV}} =  0.21 \pm 0.03
 $$
 
 \noindent 
The error in the electronic gain ratio is dominated by the uncertainty of the calibration charges. Since the QE curves of the Si and InGaAs PDs are rather flat in the region of our interest, the errors on $\overline{\epsilon_{UV}}$ and $\overline{\epsilon_{IR}}$ are small and can be neglected.

The value of $Y_{UV}$ has been calculated from the measurements available in literature,  performed in the wavelength region from 300 to about $430~nm$ at atmospheric pressure and room temperature. The data considered by us and their references are listed in table \ref{tab:yield}.  Regarding the value in \cite{Nagano2004}, the authors reported the yield as number of photons per unit length. To excite the gas, they used a $^{90}Sr$ $\beta^-$ source, which emits electrons with average energy of $0.85~MeV$. Such an energy deposit is equivalent to $0.2059~MeV/m$ \cite{Lefeuvre}. Then we can calculate the yield per deposited energy as reported in the table.

\begin{table*}[hp]
\caption{Light yields reported in literature and used to calculate $Y_{UV}$.}
\label{tab:yield}
  \begin{center}
   \begin{tabular}{c c c } \hline \hline

wavelength range (nm) & UV light  yield (ph/MeV) & reference \\ \hline \hline
$300-428$ & $19.67 \pm 0.68$ & \cite{Nagano2004}, see text  \\
$300-430$ & $20.38 \pm 0.98$ & \cite{Lefeuvre} \\
$300-420$ & $20.8 \pm 1.6$ & \cite{Abbasi2008}\\ 
$290-440$ & $17.6 \pm 2.3$ & \cite{colin2007} \\ \hline \hline
\end{tabular}
\end{center}
\end{table*}

We calculated the weighted average, which yields $Y_{UV} = 19.88 \pm 0.51 ~ph/MeV$.
Therefore, our estimate of the NIR light yield $Y_{IR}$ is

$$
Y_{IR} = 4.17 \pm 0.53~~ph/MeV
$$

\section{Discussion}

The light yield of dry air in the NIR is about $1/5$ of the UV yield. The QE of the InGaAs semiconductor is about $4$ times higher than the UV photocathode of a photomultiplier, so the total number of photoelectrons is approximately the same in the two cases. However, from a practical point of view, when compared to photomultipliers, operation with InGaAs photodiodes poses several drawbacks, such as:

- small dimension (area $< 1~cm^2$);

- no multiplication;

- need of a low noise electronics.

\noindent
Therefore, detection of cosmic showers with InGaAs PDs is nowadays not compelling.

The situation could be different if we adopt Si semiconductor, whose QE can extend down to the neighborhood of $1100~nm$, being limited by the $1.1eV$ band-gap of the ordinary bulk silicon. Moreover, photodiodes can be produced with an internal gain, obtaining Avalanche Photodiodes (APDs), with gain  in the range $10^2\div10^4$ and surface area larger than $1~cm^2$ \cite{rmd}. However, limiting the wavelength range would decrease the light output. The amount of NIR light with $\lambda \le 1100~nm$ is about $35\%$ of the total light yield in the $800\div1650~nm$ range. Therefore, Si APD in the NIR is about a factor of 10 less sensitive with respect to the UV light, at parity of detecting surface area.

So far we have not mentioned the main argument in favor of the NIR fluorescence, namely, the transparency of the atmosphere. The UV light produced by a shower $20~km$ far away from the detector is attenuated by the atmosphere by about a factor of 10. Then, since atmosphere is transparent in the NIR, detection of the NIR fluorescence with a Si APD is, in this case, equivalent to the detection of the UV light with photomultipliers.

For the general case, it is likely that the detection of UHECRs through the NIR fluorescence with Si APDs is not much disadvantaged with respect to the UV fluorescence with photomultipliers. The principal pro for the physics of UHECRs would be the increase of the event rate, which scales as $\Lambda^2$.

On the opposite side, however, it must be demonstrated that the large background noise of the atmosphere in the NIR, which is higher than in the UV \cite{background}, can be handled and does not introduce other limitations.

The technological research on Si APDs towards the enhancement of the response in the NIR is progressing and has already reached remarkable results, not yet transferred on commercial devices. For example, several groups have obtained APDs with QE above $50\%$ for $\lambda \gtrsim 1\mu m$ by laser treatment of the Si surface \cite{yamamoto,myers,wu}. Others used inclusions of SiGe absorbing layers  to enhance the QE well beyond the Si cutoff wavelength \cite{loudon}. It is therefore thinkable that in the near future the Si APDs are much more performing than the actual ones.

For sake of completeness, we have also to mention the possibility to use NIR photomultipliers, whose sensitivity now extends down to $1700~nm$ \cite{hamamatsu}. They have a small active area (smaller than the InGaAs PD) and a QE  $\leq 1\%$. Therefore operation with such photomultipliers seems not competitive with respect to large area Si APDs.

\section{Conclusions}

We have measured for the first time the fluorescence light of the air in the near infrared region. The NIR photon yield is 1/5 of the UV yield. We propose such fluorescence light as a new way to detect UHECRs. Since the atmosphere is more transparent in the NIR than in the UV region, there is reasonable expectation that the new method could lead to an increase of the UHECR observable rate.
This paper is the first step of a research program which aims to this goal.

\section{Acknowledgment}
The authors are grateful to G.Carugno and A.F.Borghesani for their support and  for useful discussions.


\begin{thebibliography}{00}


\bibitem{greisen}
K.Greisen, Ann. Rev. Nucl. Sci. 10 (1960), 63

\bibitem{delvaille}
J.Delvaille, F.Kendziorski and K.Greisen, J. Phys. Soc. Japan  17 (Suppl A 3) (1962), 76 

\bibitem{suga}
K.Suga, Proc. 5th Interamerican Seminar on Cosmic Rays, La Paz, Bolivia (1962), 2 p. XLIX

\bibitem{chudakov}
A.E.Chudakov,  Proc. 5th Interamerican Seminar on Cosmic Rays, La Paz, Bolivia (1962), 2 p. XLIX

\bibitem{hireswww}
http://www.cosmic-ray.org/

\bibitem{flyseye}
R.M.Baltrusaitis et al., Nucl. Instrum. and Meth. A 240 (1985), 410-418

\bibitem{augerwww}
http://www.auger.org/

\bibitem{TAwww}
http://www.telescopearray.org/

\bibitem{AUGER2010}
The Pierre Auger coll., J.Abraham et al., Astr. Phys. 33 (2010), 108-129

\bibitem{manducabell}
A.Manduca and R.A.Bell, Publ. Astron. Soc. Pacif., 91 (1979), 848

\bibitem{n2spectroscopy}
A.Lofthus and P.H.Krupenie, J. Phys. Chem. Ref. Data 1 (1977), 113-307

\bibitem{davidson}
G.Davidson and R.O'Neil, J. Chem. Phys. 41 (1964), 3946

\bibitem{kimball}
Kimball Physics Inc., Wilton NH, USA

\bibitem{diamondmaterials}
Applied Diamond Inc., Wilmington DE, USA

\bibitem{sipd}
mod. S1337-1010BQ by Hamamatu Photonics, Japan

\bibitem{ingaaspd}
mod. J22TE2-66C-R05M-1.7 by Teledyne Judson Technologies, USA

\bibitem{QAUA1}
C.Bacci et al., Nucl. Instrum. and Meth. A 273 (1988), 321-325

C.Bacci et al., Nucl. Instrum. and Meth. A 279 (1989), 169-179

\bibitem{chamberlain}
J.Chamberlain, {\it The principles of interferometric spectroscopy}, John Wiley and Sons, 1979

\bibitem{griffith}
P.R.Griffiths, J.A.de Haseth, {\it Fourier Transform Infrared Spectrometry}, John 
Wiley and Sons, 2007 

\bibitem{nist}
Yu.Ralchenko, A.E.Kramida, J.Reader, and NIST ASD Team (2008). NIST Atomic Spectra Database (version 3.1.5), [Online]. Available: http://physics.nist.gov/asd3 . National Institute of Standards and Technology, Gaithersburg, MD. 

\bibitem{krupenie_o2}
P.H.Krupenie, J. Phys. Chem. Ref. Data, 1 (1972), 423-534

\bibitem{spettrouv}
M.Ave et al., Nucl. Instrum. and Meth. A 597 (2008), 41-45

\bibitem{Nagano2004}
M.Nagano et al., Astr. Phys. 22 (2004), 235-248

\bibitem{Lefeuvre}
G.Lefeuvre et al., Nucl. Instrum. and Meth. A 578 (2007), 78-87

\bibitem{Abbasi2008}
R.Abbasi et al., Astr. Phys. 29 (2008), 77-86

\bibitem{colin2007}
P.Colin et al, Astr. Phys. 27 (2007), 317-325

\bibitem{rmd}
Radiation Monitoring Devices Inc., Watertown MA, USA

\bibitem{background}
Ch.Leinert at al., Astron. Astrophys. Suppl. Ser. 127 (1998), 1-99

\bibitem{yamamoto}
K.Yamamoto et al., Nucl. Instrum. and Meth. A (2010), doi:10.1016/j.nima.2010.03.128

\bibitem{myers}
R.A.Myers et al., Appl. Opt. 45 (2006), 8825-8831

\bibitem{wu}
C.Wu et al., Appl. Phys. Lett. 78 (2001), 1850-1852

\bibitem{loudon}
A.Y.Loudon et al., Opt. Lett. 27 (2002), 219-221

\bibitem{hamamatsu}
mod.R5509-72 by Hamamatsu Photonics, Japan.

\end{thebibliography}
\end{document}